\documentclass[preprint,aps,prb]{revtex4}   
\include{amsmath}
\include{epsfx}
\usepackage{graphicx}
\begin{document}

\title{{\bf  Proteins in a shear flow}}

\author{{\bf P. Szymczak$^1$ and Marek Cieplak$^2$}}

\affiliation{
$^1$Institute of Theoretical Physics, Warsaw University,
ul. Ho\.za 69, 00-681 Warsaw, Poland\\
$^2$Institute of Physics, Polish Academy of Sciences,
Al. Lotnik\'ow 32/46, 02-668 Warsaw, Poland}

\begin{abstract}
{The conformational dynamics of a single protein molecule in a shear flow is
investigated using Brownian dynamics simulations. A structure-based coarse grained model
of a protein is used. We consider two proteins, ubiquitin and integrin, and find that
at moderate shear rates they unfold through a sequence of metastable states -- a
pattern which is distinct from a smooth unraveling found in homopolymers.
Full unfolding occurs only at very large shear rates. Furthermore, the hydrodynamic
interactions between the amino acids are shown to hinder the shear flow unfolding.
The characteristics of the unfolding process depend on whether a protein is
anchored or not, and if it is, on the choice of an anchoring point.
}

\end{abstract}

\maketitle

\vskip 40pt
\noindent {\bf
Keywords: protein stretching, protein folding, manipulation of
proteins, Go model,
molecular dynamics, integrin, ubiquitin}

\noindent {PACS numbers: 82.37.Rs, 87.14.Ee, 87.15.-v}

\vspace*{1cm}

\section{Introduction}
Mechanically induced conformational changes in biomolecules can be accomplished
experimentally in many ways. One of them is by using a pulling device,
such as an atomic force microscope, another is by involving fluid flow.
The pulling device can be used in two basic modes: at a constant pulling velocity
or at a constant force. The latter mode results in a nearly homogeneous
tension along the backbone of a biomolecule and in an essentially two-state
unfolding behavior in simple proteins such as ubiquitin \cite{Schlierf2004}.
The flow induced stretching is very different in this respect since - even in the
case of the uniform flow - the tension along the backbone increases as
one moves from the free end to the tethered end \cite{Marko1995,Rzehak2000}.
This effect was predicted \cite{Szymczak-Cieplak:2006}  to lead to emergence
of many partially unfolded metastable conformations which arise when one ramps-up
speeds of flow.

Experiments on flow-generated stretching of biomolecules have been performed either
by using the direct bulk flow \cite{
Perkins1995,Larson1997,Smith1999,Doyle2000,Teixeira2005,Schroeder2005a,
Schroeder2005b,Shaqfeh2005} or by using the meniscus forces, for example in molecular
combing \cite{Strick:2003,Bensimon:1996,Austin:1998,Inganas:2005} or the combination of the
viscous and meniscus forces, as it is the
case in spin-stretching \cite{Yokota:1999,Kim:2007} technique.
The experiments have been usually performed on the long DNA chains.
However, in the case of proteins a full unfolding of a molecule requires flow velocities
that are
three orders of magnitude larger \cite{Szymczak-Cieplak:2006} than those needed for the DNA
unfolding because of the much smaller hydrodynamic radii and larger intramolecular forces
involved.
Nevertheless, stretching proteins at smaller flow velocities could still produce interesting
effects, since it usually results in the formation of partially unfolded intermediates.
However, in contrast to AFM force spectroscopy, there are still no experiments on the
protein unfolding in the flow on the single-molecule level. Instead, flow denaturation
experiments were carried out on a bulk collection of molecules, usually subject to the
shearing forces \cite{Charm1970,Thomas1979,Thomas1979b,Maa1996,Hagen2006}.
A notable exception are the studies on the von Willebrand factor (vWf), a large
poly-protein found in the blood plasma \cite{Katz2006,Schneider2007,Siedlecki1996}. However,
although the existence of shear-induced conformational transition in that system is well
documented, it does not seem to involve the unfolding of individual domains of the protein.
Instead, a conformational transition of the whole multi-unit chain takes place, from a
compact globular state to the elongated fiber-like conformation.

The experimental results quoted here seem to suggest that the shear rates needed to unfold a
small protein are extraordinarily high. For example Jaspe and Hagen \cite{Hagen2006}
tried to unfold horse cytochrome with the shear rates as high as $10^5$ Hz
but observed no evidence of the shear destabilization of the folded state of the protein.
They also gave a simple estimate that in order to unfold a protein, the shear rate should be
characterized by the Weissenberg number, $\text{Wi}=s \tau_{fold}$ of the order of $10^3$.
Here $\tau_{fold}$ is the folding time of a protein and $s$ is the
shear rate. We confirm that prediction through molecular dynamics simulations of a protein
using a coarse-grained model. Additionally, we show that stretching of proteins by shear
flow proceeds differently from that of homopolymers.

The simulations were conducted for ubiquitin (1ubq) and integrin (1ido).
Both have been the subject of our previous investigations on protein stretching in uniform
flow. In particular, integrin was shown to possess a surprisingly rich spectrum of
metastable states when
stretched by the flow  \cite{Szymczak-Cieplak:2006} which makes it a perfect system in which
to investigate the effects
of the shear flow.

\section{The model}

Coarse-grained models of proteins allow one to access time scales which are orders of
magnitude longer than those available in all-atom simulations. Among the coarse-grained
models, the Go-like implementations \cite{Abe1981,Takada1999} link the properties of a
protein directly to its native geometry and are probably the easiest to use. We follow the
implementation presented in references \cite{
Hoang2000b,Hoang2000,Cieplak2002,Cieplak2003,Sulkowska-Cieplak:2007}. The protein
is represented by a
chain of its C$^{\alpha}$ atoms. The successive C$^{\alpha}$ atoms along the backbone are
tethered by harmonic potentials with a minimum at 3.8 {\AA}.
The effective interactions between the residues are split into two classes: native and non-
native. This determination is made by checking for native overlaps between the enlarged van
der Waals surfaces of the amino acids as proposed in reference \cite{Tsai1999}. The amino
acids, $i$ and $j$, that do overlap in this sense are endowed with the effective Lennard-
Jones potential $V_{ij} = 4\epsilon \left[ \left( \frac{\sigma_{ij}}{r_{ij}}
\right)^{12}-\left(\frac{\sigma_{ij}}{r_{ij}}\right)^6\right]$.
The length parameters $\sigma _{ij}$ are chosen so that the potential
minima correspond, pair-by-pair, to the experimentally established
native distances between the respective amino acids in the native state.
 Non-native contacts are represented by hard core repulsion in order to prevent
entanglements. Another term in the Hamiltonian imposes local stiffness on the
backbone. This can be accomplished either by introducing biases in the
bond and dihedral angles \cite{Clementi2000} or by favoring native senses of local
chiralities \cite{Kwiecinska2005}. Here, we choose the latter.

The protein was subject to a simple shear flow of the form
\begin{equation}
v_x = s y, \ \ \  v_y=v_z=0
\end{equation}
which may also be written as ${\bf v}= {\bf K} \cdot {\bf r}$
where $K$ is the velocity gradient matrix, in this case given by
\begin{equation}
K=\left( \begin{array}{ccc}
0 & s & 0 \\
0 & 0 & 0 \\
0 & 0 & 0 \end{array} \right)
\end{equation}
In our previous studies on uniform flow unfolding \cite{Szymczak-Cieplak:2006} we used the
Langevin dynamics method to track the evolution of the system. Here, however, we use the
Brownian
dynamics scheme since it allows for a straightforward incorporation of hydrodynamic
interactions (HI). Without the hydrodynamic interactions, the two
schemes (Langevin and Brownian dynamics) give the same results, since on the time scales
characteristic for protein unfolding, the motion is overdamped and inertia effects are
negligible, as confirmed in Ref. \cite{Szymczak-Cieplak:2007}.

In the Brownian dynamics algorithm \cite{Ermak-McCammon:1978},
the displacement of the $i$'th amino acid during the timestep $\Delta t$ given by
\begin{equation}
 \Delta {\bf r}_i = {\bf K} \cdot {\bf r}_i \Delta t +  \sum_j \bigl( \nabla_j \cdot {\bf D}
_{ij} \bigr) \Delta t
+ \frac{1}{k_B T} \sum_j {\bf D}_{ij} \cdot {\bf F}_j \Delta t + {\bf C}_i : {\bf K} \Delta
t
+ {\bf B}_i,
\label{nar1}
\end{equation}
Here ${\bf r}_i$ is the position of i-th amino acid, ${\bf F}_i$ is the total interparticle
force acting on
it, ${\bf D}$ is the diffusion tensor. Note that both ${\bf F}$ and ${\bf D}$ are
configuration-dependent. Next, ${\bf B}$ is a random displacement given
by a Gaussian distribution with an average value of zero and covariance
obeying
 \begin{equation}
<{\bf B}_i {\bf B}_j> = 2 {\bf D}_{ij} \Delta t.
\label{Gauss}
\end{equation}
Finally, ${\bf C}$ is the third rank shear disturbance tensor \cite{Dhont1996} representing
the
effect
of interparticle hydrodynamic forces on the shear-induced particle motion
\cite{Mazur1982,Dickinson1994}. Most of our simulations were conducted without taking
hydrodynamic interactions into account (free draining
model). In this case, the diffusion tensor is diagonal
\begin{equation}
{\bf D}_{ij}=\frac{k_B T}{6 \pi \eta a} {\bf I} \delta_{ij}
\label{nohi}
\end{equation}
and the disturbance matrix vanishes. At the end of the paper, we discuss the influence of
the hydrodynamic interactions on the mean extension at various shear rates.
In that case, we use the Rotne, Prager and Yamakawa
\cite{Rotne1969,Yamakawa1970}
approximation of the diffusion tensor, with the nondiagonal terms of ${\bf D}_{ij}$ given by
\begin{equation}
{\bf D}_{ij}=\displaystyle \frac{k_B T}{8 \pi \eta r_{ij}} \left\{\begin{array}{cl}
      \displaystyle \left[ \left( 1+\frac{2a^2}{3 r_{ij}^2} \right) {\bf I} + \left(1-
\frac
{2a^2}{r_{ij}^2}\right)
{\bf \hat{r}}_{ij} {\bf \hat{r}}_{ij} \right], & r_{ij} \geq 2a\ \\
\\
\displaystyle \frac{r_{ij}}{2a} \left[ \left( \frac{8}{3}-\frac{3r_{ij}}{4a} \right) {\bf I}
+
\frac
{r_{ij}}{4a}
{\bf \hat{r}}_{ij} {\bf \hat{r}}_{ij} \right], & r_{ij}  < 2a
         \end{array}\right.
\label{rp}
\end{equation}
where ${\bf r}_{ij}={\bf r}_{j}-{\bf r}_{i}$  and $a$ represents the hydrodynamic radius of
a bead.

\begin{figure}
\vspace{1cm}
\includegraphics[width=8cm, viewport=30 100 500 700]{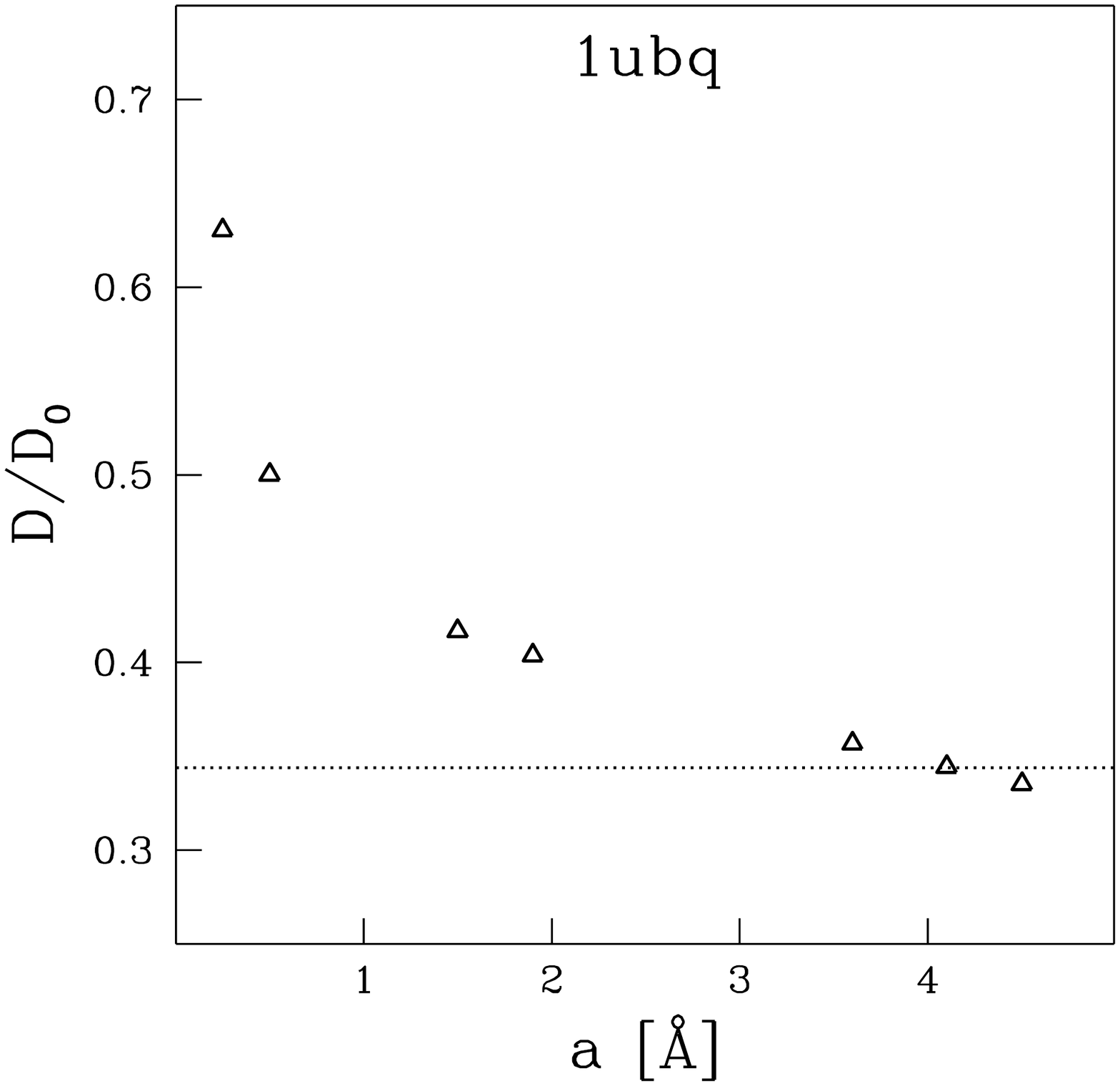}
\caption{ }
\label{diffusion}
\end{figure}

When taking the hydrodynamic interactions into account, the choice of a hydrodynamic radius,
$a$, is a crucial element in the model. One of the ways of tuning this parameter is to
compare the translational diffusion coefficient $D$, of a
protein in a numerical model, to the one measured in experiment. Fig.~\ref{diffusion} shows
the dependence of $D'=D/D_0$ on the hydrodynamic radius. Here the diffusion coefficients $D$
are normalized by $D_0={kT}/{6\pi\eta R}$ - the diffusion coefficient of the bead of
radius $R$=5 {\AA}. In water at $T=300 K$, with viscosity of $\eta =0.89 \cdot 10^{-2}$
Poise, one gets $D_0=4.93 \cdot 10^{-6} cm^2/s$.
The experiment \cite{Dingley:1997} gives the diffusion coefficient of ubiquitin
$D=1.7 \cdot 10^{-6} cm^2/s$, or, equivalently,
$D'=0.345$. As can be seen from the data, the agreement with experiment is obtained for
$a \approx 4.1$. This value agrees with earlier estimates by de la Torre and Antosiewicz
\cite{Torre:1981,Antosiewicz:1989,Hellweg:1997,Torre:2000,Banachowicz:2000}. However, since
the distance between the successive C$^{\alpha}$ atoms along the protein backbone is $3.8$ {
\AA}, some of
the beads representing amino acids overlap. This reflects the fact that: 1) the interior of
the
protein is densely packed, 2) the side chains of amino acids are usually longer than $3.8$
\AA , and 3) the protein is covered by the  hydration layer of tightly bound water
molecules.
Although the Rotne-Prager tensor is positive definite also for overlapping beads, its
physical meaning for such configurations is problematic \cite{Torre:2000}. The overlapping
bead models are successful in predicting the diffusion coefficients of the proteins \cite{
Torre:1981,Antosiewicz:1989}, however the question whether they correctly reproduce the
dynamic effects of hydrodynamic interactions during large-scale conformational motions in
macromolecules is still open.
To investigate the influence of the hydrodynamic radius on the dynamics of the protein in
the flow, we compare the results obtained for $a=4.1$ with those for a smaller value of the
hydrodynamic radius, $a=1.5$ {\AA}. The analysis of Fig.~\ref{diffusion} shows that the
choice of
$a=1.5$ {\AA} overestimates the diffusion coefficients by 20\% only, but has the advantage
of
not leading to overlapping configurations of the spheres with the radii of $a$.
Such a value of $a$ was also used
in our previous studies on the influence of hydrodynamic interactions on protein unfolding
\cite{Szymczak-Cieplak:2007} whereas Baumketner and Hiwatari \cite{Baumketner:2002} use
$a=1.77$ in their
investigations on the influence of HI on the protein folding process.

A natural time scale in the simulations, $\tau$, is set by the time it takes for the
amino acid (i.e. the object of the radius $a=4.1$ {\AA}) to diffuse the distance of its
radius, $a$. Again, in water at $T=300 K$, one gets $\tau =0.05 ns$. This time scale is used
as a time unit in the data reported, irrespectively of the actual value of hydrodynamic
radius used ($4.1$ or $1.5$ {\AA}).

The folding time for ubiquitin,
as calculated according to the scheme described e.g.
in~\cite{Hoang2000} in the free-draining case (with $a=4.1$ \AA) is equal to
$\tau_{fold} \approx 1100 \tau$ whereas for integrin $\tau_{fold} \approx 10000 \tau$. When
the
hydrodynamic interactions are taken into account ubiquitin folding time becomes
$\tau_{fold} \approx 370 \tau$ for $a=4.1$ and $\tau_{fold} \approx 150 \tau$
for $a=1.5$. The latter difference is caused mainly by the differences in the single
amino acid diffusion coefficients in both cases; the ratio of the diffusion coefficients
$(4.1/1.5) \approx 2.7$ is only about 10\% larger than the corresponding ratio of folding
times, $(370/150) \approx 2.5$.

The initial conformation of the protein corresponds to the native structure. During the
simulations either one of the termini of the protein is anchored or both ends are free.
In both cases, the fluid in which the protein is immersed is unbounded in all directions
The end-to-end extension, the orientation angle and the root-mean-square deviation (RMSD)
from the native structure are
then recorded as a function of time. The total length of the simulation time for each
trajectory was $t=400 000 \tau$ for the free-draining case (for both ubiquitin and integrin)
and $t=10 000 \tau$ for the simulations with hydrodynamic interactions (for ubiquitin).
The non-dimensional flow strength is characterized by the Weissenberg number,
$\text{Wi}=s \tau_0$,
where, following Jaspe and Hagen \cite{Hagen2006}, we take $\tau_0$ to be equal to the
folding time of
the protein, $\tau_{fold}$.

\section{Results}

As already noted by Lumley \cite{Lumley1969} and de Gennes \cite{Gennes1974}, a notable
feature of the shear flow is that it is a combination of the elongational and rotational
components of equal magnitudes. In such a marginal case the polymer chain does not attain a
stable stretched configuration. Instead, it undergoes a tumbling motion, a series of
subsequent stretching and coiling events with frequent changes in the orientation of the
chain with respect to the shear axis \cite{Smith1999,Doyle2000,Teixeira2005,Schroeder2005a}.
While the elongational component of the flow is stretching the molecule, its rotational
component aligns it along the shear axis, leading to the collapse of the chain due to the
decreased hydrodynamic drag. An important role in this dynamics is played by the Brownian
fluctuations, which cause the chain segments to cross the streamlines into the regions of
higher or lower flow which results in further stretching or collapse of the chain
respectively. In particular, the fluctuations may tip the polymer in such a way that its two
ends lie in the regions of opposite flow direction, which results in a tumbling event,
in which one polymer end moves over the other.

We show that the tumbling dynamics are also seen in the case of protein stretching by a
shear flow. As an
example, Figs. 2 and 3 show the time series of configurations of integrin in shear flow
for both tethered and free protein. However, the presence of a complex network of bonds
between amino acids in a
protein results in a number of important differences between the homopolymer and protein
unfolding. In particular, the extension of the protein in the uniform flow is not a
continuous function of the flow rate. Instead, as the flow velocity is increased, the
protein undergoes a number of rapid transitions to the successive metastable states.
Each of those transitions is accompanied by the breaking of a particular group of bonds and
unzipping of subsequent structures from the bulk of the protein. As an example,
Fig.~4 shows the set of intermediates arising during unfolding of the
integrin molecule in a uniform flow.

The presence of intermediate states is also observed in the case of the shear flow.
However, in a shear flow those states are never long-lived, even a small thermal fluctuation
may move the protein to the region of smaller flow and the molecule collapses.
Nevertheless, some fingerprints of the underlying discrete set of intermediate states are
present even in the histogram of end-to-end distances, which shows the maxima corresponding
to the metastable states observed in the uniform flow stretching. As an example,
Fig.~\ref{fig3} shows the end-to-end length distribution for the integrin molecule tethered
by the $C$ terminus. Such a peak structure in the histogram is observed for medium shear
rates only;
lower shear rates are not strong enough for the intermediates to overcome the free energy
barrier needed to partially unfold the chain whereas in the case of high shear rates the
tumbling rate increases, the protein spends even less time in the stationary
conformations and thus the histogram no longer shows the intermediates. Fig.~\ref{fig4}
gives
an example of the end-to-end length distribution for the 50~\% higher shear rate than that
in
Fig.~\ref{fig3}. As it is seen, the peak structure is now almost impossible to discern.

It is worth noting that in the case of homopolymer in a shear flow, the respective
probability distributions have a much simpler structure. As an example, Fig.~\ref{fig5}
shows
the RMSD histogram for a simple helix (of 48 residues) in a shear flow. Also the histograms
of the chain extension in DNA experiments are usually much smoother, with one or two maxima
only \cite{Schroeder2005b}.

The characteristics of uniform flow unfolding are asymmetric with respect
to the protein anchoring \cite{Szymczak-Cieplak:2006}. For example, integrin unfolds more
easily when tethered by the C terminus. Also, the C terminus tethering leads to a much
richer spectrum of intermediates than the N terminus tethering. As mentioned in the
Introduction, this asymmetry is caused by the fact that the tension along the protein
backbone increases from the free end towards the tethered one when the molecule is placed in
the flow. Such an asymmetry can also be seen in the shear flow unfolding. In particular,
Fig.~\ref{extension} shows the relative extension ($L/L_0$) of the integrin molecule as a
function of the Weissenberg number for different tethering points. Here the average end-to-
end distance of the molecule (L) is normalized by the maximum extension length
$L_0=(N-1) \times 3.8$ {\AA}, where $N$ is the number of amino acids.
It is seen that the C terminus tethering allows unfolding at a lower shear rate,
which is consistent with the uniform flow stretching results. The critical
shear rate, at which the unfolding events begin to take place is shifted from
$\text{Wi}=600$ in
the case of the C terminus tethering to about $\text{Wi}=1300$ for the N terminus tethering.
Finally, when both ends of a protein are free, the critical shear rate is the same as in the
$N$ tethering case ($\text{Wi}=1300$), but the shape of the dependence of $L/L_0$ on Wi is
slightly
different from that observed for the $N$ tethering.

As noted by Smith \cite{Smith1999} and Doyle \cite{Doyle2000}, the power spectral density of
the end-to-end polymer extension in a shear flow shows no distinct peaks which indicates
that
no periodicity is present in the evolution of this variable. However, it was subsequently
reported \cite{Schroeder2005a,Delgado2006} that there are other variables characterizing the
polymer motion which show the periodic behavior. In particular, a well defined
characteristic tumbling frequency is revealed in the spectrum of a polymer orientation angle
$\theta$ \cite{Schroeder2005a}. Analogous spectrum for the protein is presented in
Fig.~\ref{spectrum}. Two peaks can be identified in the spectrum, corresponding to two
characteristic
tumbling frequencies. The higher frequency is the one associated with the rotational
component of the shear flow, $f_1=\frac{s}{4 \pi}$ (a sphere immersed in the shear field
rotates with the frequency $f_1$ \cite{Dhont1996}). On the other hand, the lower frequency,
$f_2$, corresponds to the stretching-collapse cycle and scales sublinearly with the flow
rate, similarly to what was reported in DNA experiments \cite{Schroeder2005a}.
 Both dynamical behaviors are observed in Fig.~\ref{angle} which
shows the time trace of the protein orientation angle and the associated evolution of the
RMSD. In the high frequency regime, the protein is closely packed, with small values of RMSD
away form the native structure and it essentially behaves like a sphere rotating in a shear
flow. From
time to time, a sudden unfolding event takes place - the protein ceases to rotate and its
RMSD rapidly increases. For larger shear rates, the unfolding events
are more
frequent and the periods of free rotation in the globular state - shorter, as seen in Fig~
\ref{anglestrong}. Fig.~\ref{freq} shows the tumbling frequencies, $f_1$ and $f_2$
as the function of Wi. Note that there is a relatively narrow shear rate range for which
both
frequencies are visible in the spectrum. For lower shear rates the protein hardly ever
unfolds hence $f_2$ is impossible to discern. On the other hand, at large shear rates the
protein is never found in a globular state for a sufficient period of time for $f_1$ to be
seen.

The inclusion of hydrodynamic interactions considerably hinders the unfolding of a molecule.
Fig.~\ref{ubi} shows the comparison of the relative extension of ubiquitin in a shear flow
calculated with and without the hydrodynamic interactions. It is seen that, when HI are
included in the model, much larger shear rates are needed for the unfolding of a molecule
and the corresponding extensions of the chain are significantly smaller.
In particular, a critical shear rate needed for the unfolding events to take place is
shifted from $\text{Wi}=250$ for the non-HI case to $\text{Wi}=600$  (HI present, $a=1.5$)
and $\text{Wi}=2000$
(HI present, $a=4.1$). This is consistent with our earlier studies on the uniform flow
unfolding \cite{Szymczak-Cieplak:2007} where it was observed that unfolding of the system
with HI requires a much larger flow speed than without. This tendency can be understood
qualitatively in terms of the so-called non-draining effect \cite{Rzehak:2000}: the residues
hidden inside the protein are shielded from the flow and thus only a small fraction of the
residues experience the full drag force of $F=-\gamma U$. In contrast, when no HI are
present, this drag force is applied to all residues. This effect is analyzed more
quantitatively e.g. in Ref.~\cite{Agrawal:1994} where, in particular, it is shown that
the hydrodynamic shielding increases with an increase in both the hydrodynamic radius and
the total number of beads in the polymer chain.

\section{Summary}

We have presented Brownian Dynamics simulation results on the conformational dynamics of
individual protein molecules in the flow. The presence of a network of bonds between amino
acids in a protein leads to a number of important differences between the homopolymer and
protein unfolding. In particular, in the case of proteins, the characteristics of the
unfolding process are shown to depend on the selection of the point of anchor.
Additionally, for moderate shear rates, a number of intermediate stages in the unfolding
may be discerned, with well defined RMSD values with respect to the native structure.
Full unfolding of the proteins was found to occur only at very high shear rates, which is
is consistent with the experimental results \cite{Hagen2006}.

\section{Acknowledgments}

This project has been supported by the
Polish Ministry of Science and Higher Education (Grant N N202 0852 33) and by the European
program IP NaPa through Warsaw University of Technology.
Fruitful discussions with Jan Antosiewicz are appreciated.

\bibliography{013741JCP}

\newpage
\centerline{FIGURE CAPTIONS}

\begin{description}

\item[Fig. 1. ]
The reduced diffusion coefficient as a function of a hydrodynamic radius of an
amino acid. The horizontal line marks the experimental value $D/D_0 \approx 0.345$.

\item[Fig. 2. ]
An example of cycle of the motion of the tethered protein motion in shear flow: integrin at
$s=0.25 \tau^{-1}$ (\text{Wi}=2000). The anchoring point (C terminus) is marked by a
circle.

\item[Fig. 3. ]
An example of cycle of the motion of a free protein in shear flow: integrin at
$s=0.4 \tau^{-1}$ (\text{Wi}=3200). For tracing purposes, half of the chain is colored red,
and another half - green.

\item[Fig. 4. ]
Integrin tethered by the C terminus. (Upper) Examples of the time evolution of the
RMSD
from the native structure in unfolding of integrin in a uniform flow for various flow rates.
The plateaus correspond to successive stationary conformations (intermediates) marked by the
letters (A-E). The snapshots of conformations A,B,D, and E are shown on the right.
(Lower) The histogram of RMSD for the integrin in a shear flow at \text{Wi}=2000. The
respective
values of RMSD corresponding to the intermediates seen in the upper panel are marked.

\item[Fig. 5. ]
Same as in the lower panel of Fig. 2 but for $s=0.4 \tau^{-1}$ (\text{Wi}=3200).

\item[Fig. 6. ]
The histogram of RMSD for the helix (48 residues) in a shear flow at $s=0.25
/\tau$.

\item[Fig. 7. ]
The relative extension of the integrin molecule as a function of the Weissenberg
number for a chain tethered by the C terminus (filled triangles), N terminus (empty
triangles) and a free chain (squares). The average end-to-end distance of the molecule (L)
is normalized by the maximum extension length $L_0=(N-1) \times 3.8$ {\AA}, where $N$ is the
number of amino acids.

\item[Fig. 8. ]
Power spectral density (psd) of the protein orientation angle for integrin tethered
by the N terminus in a shear flow at \text{Wi}=1600. Frequencies are scaled by the protein
folding time and the psd is normalized with its maximum value.

\item[Fig. 9. ]
The angle $\theta$ between the end-to-end direction of the protein and the
direction
of the flow (upper panel) and RMSD for integrin tethered by the C terminus in a shear flow
at $\text{Wi}=640$ (lower
panel).

\item[Fig. 10.]
Same as in Fig. 9 but for  $\text{Wi}=960$.

\item[Fig. 11.]
The peak frequencies, $f_1$ and $f_2$ derived from the power spectrum densities of
orientation angle as a function of the Weissenberg number for integrin tethered by the C
terminus (empty squares) and the N terminus (filled squares). The solid line corresponds to
the relation $f_1=s/4\pi$.

\item[Fig. 12. ]
The relative extension of ubiquitin anchored by the N terminus as a function of the
Weissenberg number for the model without hydrodynamic interactions (filled triangles), and
with hydrodynamic interactions for $a=1.5$ {\AA}  (open triangles) and $a=4.1$ {\AA} (
squares).

\end{description}

\newpage

\newpage

\begin{figure}
\vspace{3cm}
\includegraphics[width=16cm, viewport=0 250 550 650]{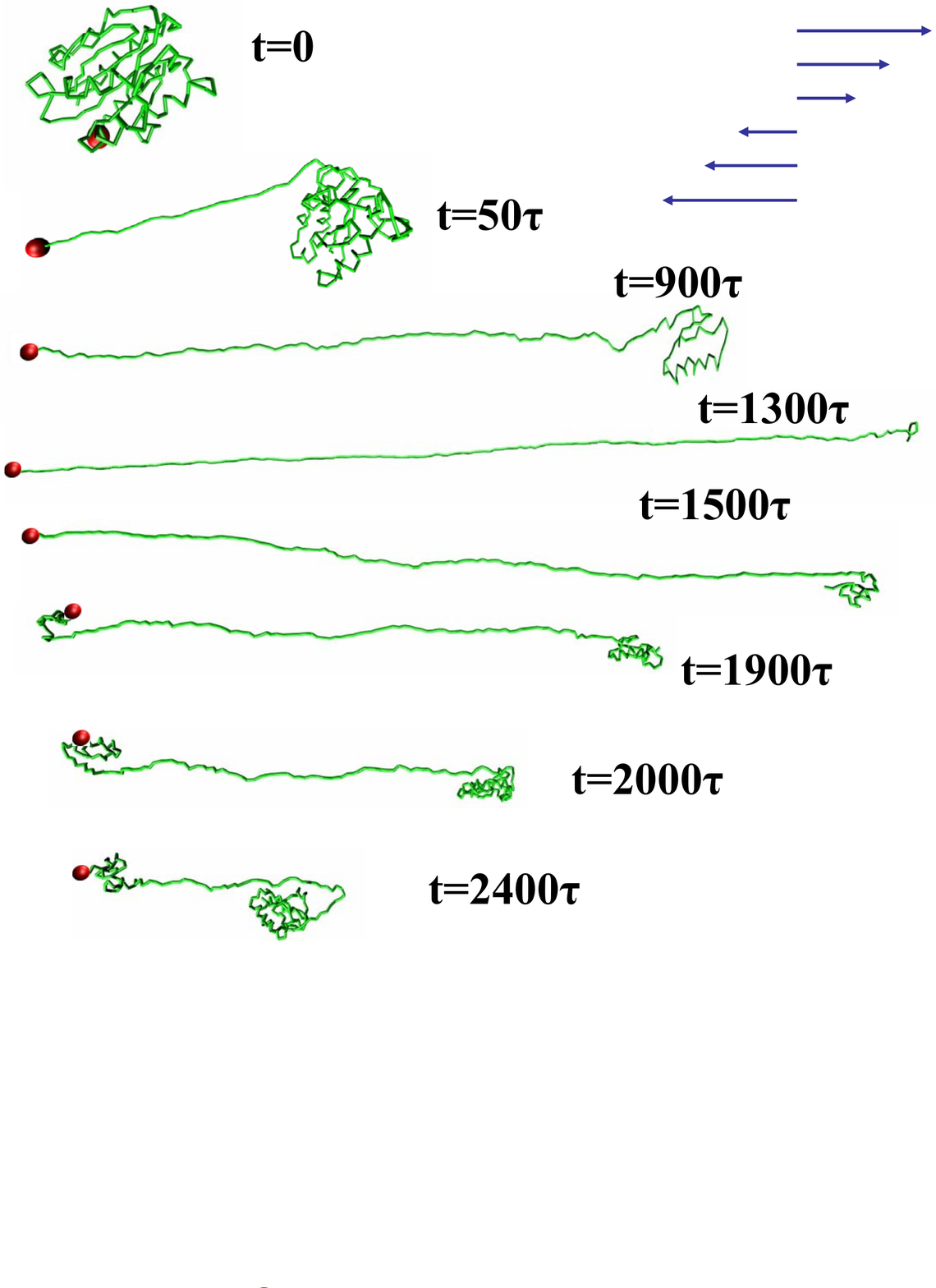}
\vspace{4cm}
\caption{ }
\label{fig1}
\end{figure}

\newpage

\begin{figure}
\vspace{2cm}
\includegraphics[width=10cm, viewport=100 150 500 500]{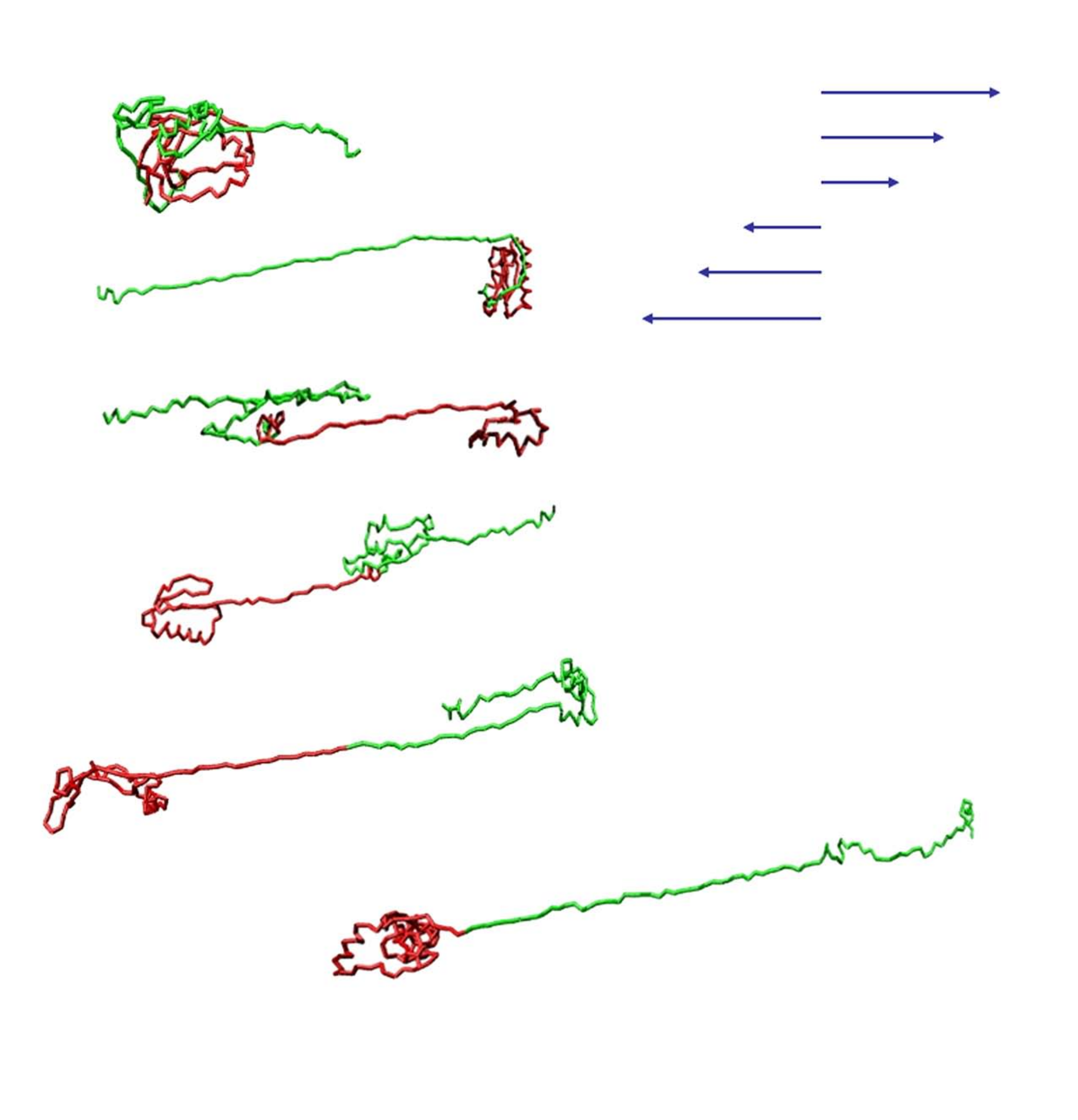}
\vspace{2cm}
\vspace{2cm}
\caption{ }
\label{fig2}
\end{figure}

\newpage

\begin{figure}
\vspace{2cm}
\includegraphics[width=16cm, viewport=0 200 500 700]{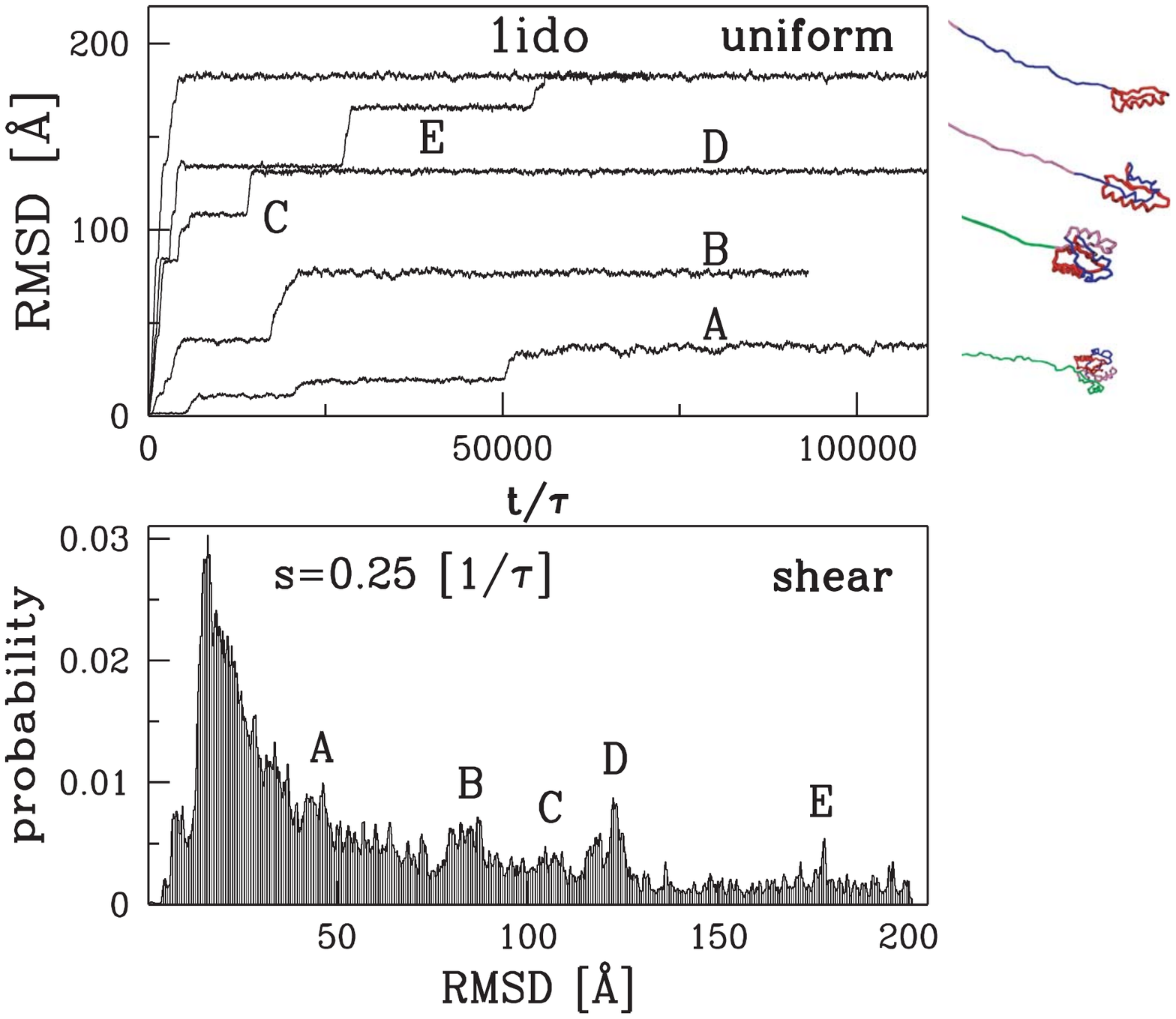}
\caption{ }
\label{fig3}
\end{figure}

\newpage

\begin{figure}
\vspace{1cm}
\includegraphics[width=10cm, viewport=30 100 500 500]{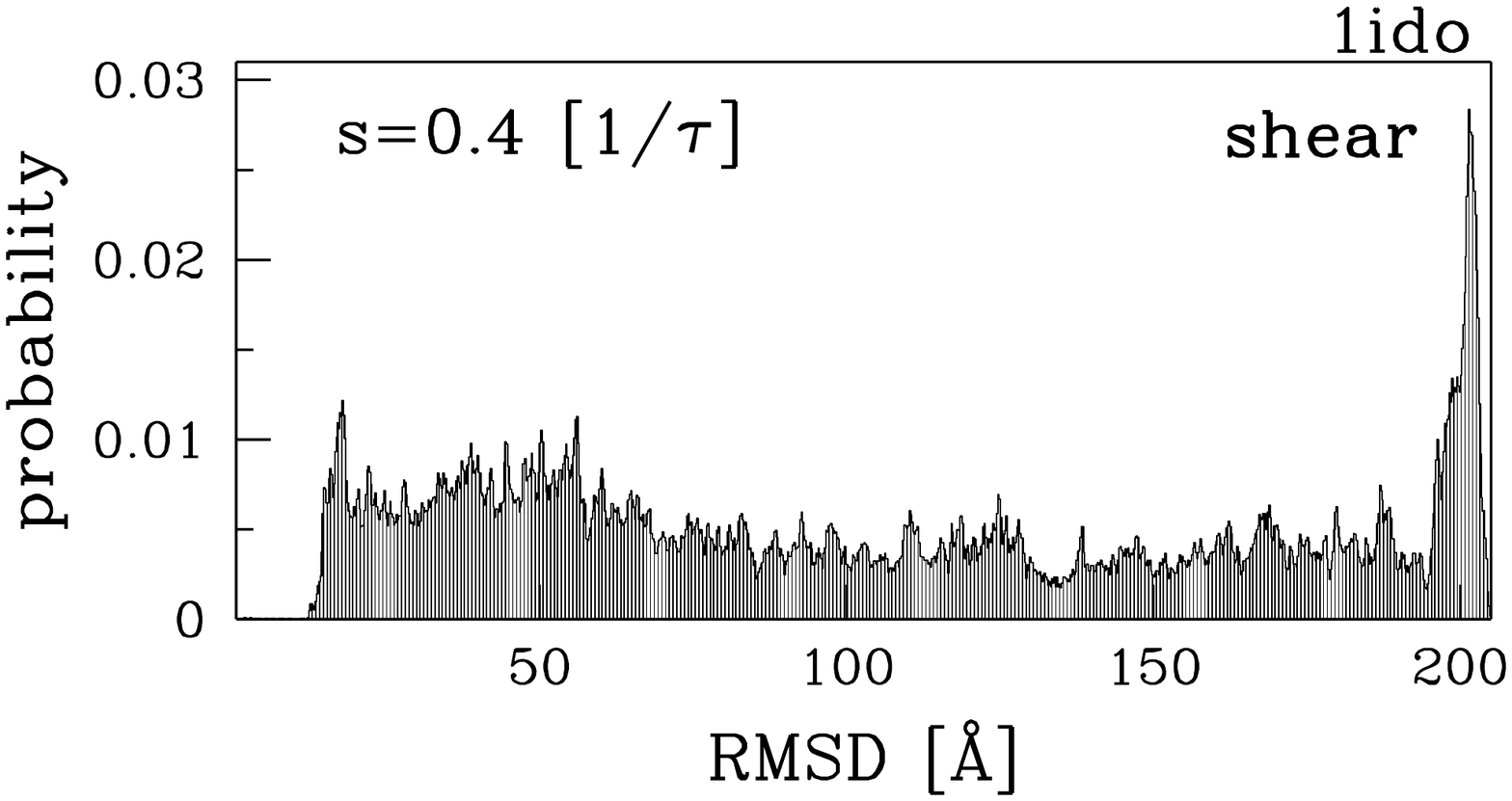}
\vspace{-1cm}
\caption{ }
\label{fig4}
\end{figure}

\newpage

\begin{figure}
\vspace{1cm}
\includegraphics[width=10cm, viewport=30 100 500 500]{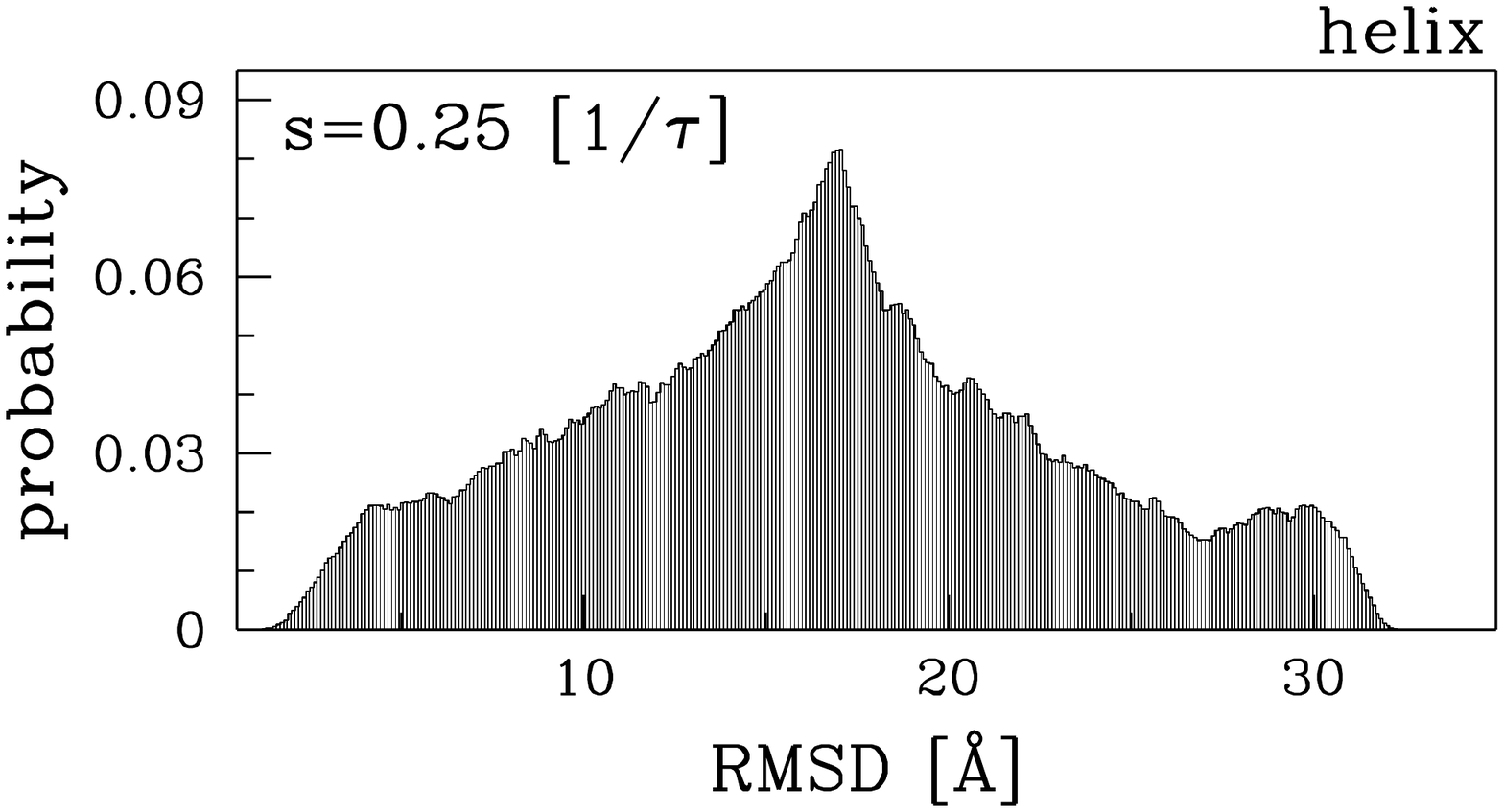}
\vspace{-1cm}
\caption{ }
\label{fig5}
\end{figure}

\newpage

\begin{figure}
\vspace{1cm}
\includegraphics[width=8cm, viewport=30 100 500 700]{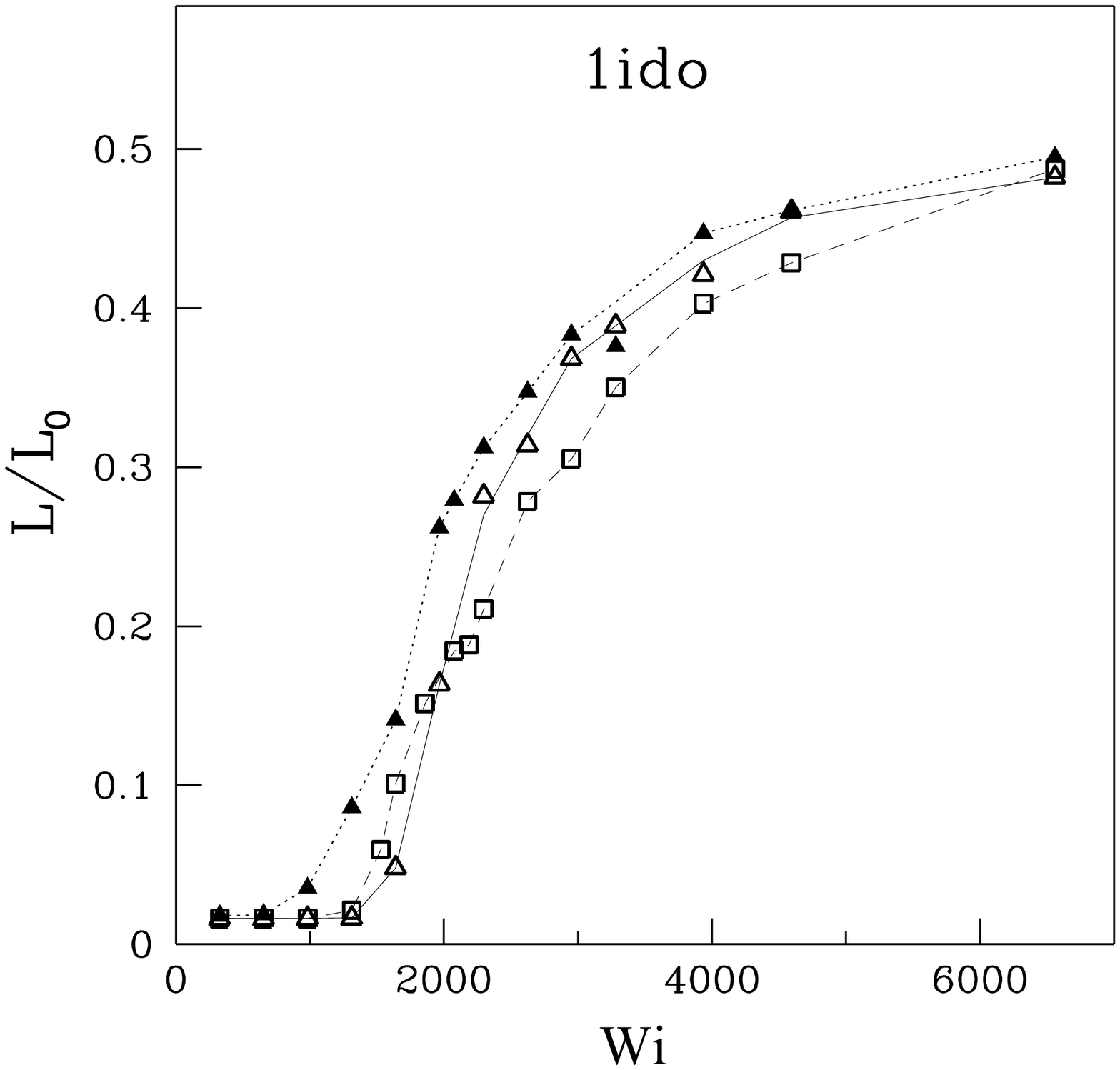}
\caption{ }
\label{extension}
\end{figure}

\newpage

\begin{figure}
\includegraphics[width=12cm, viewport=30 100 500 700]{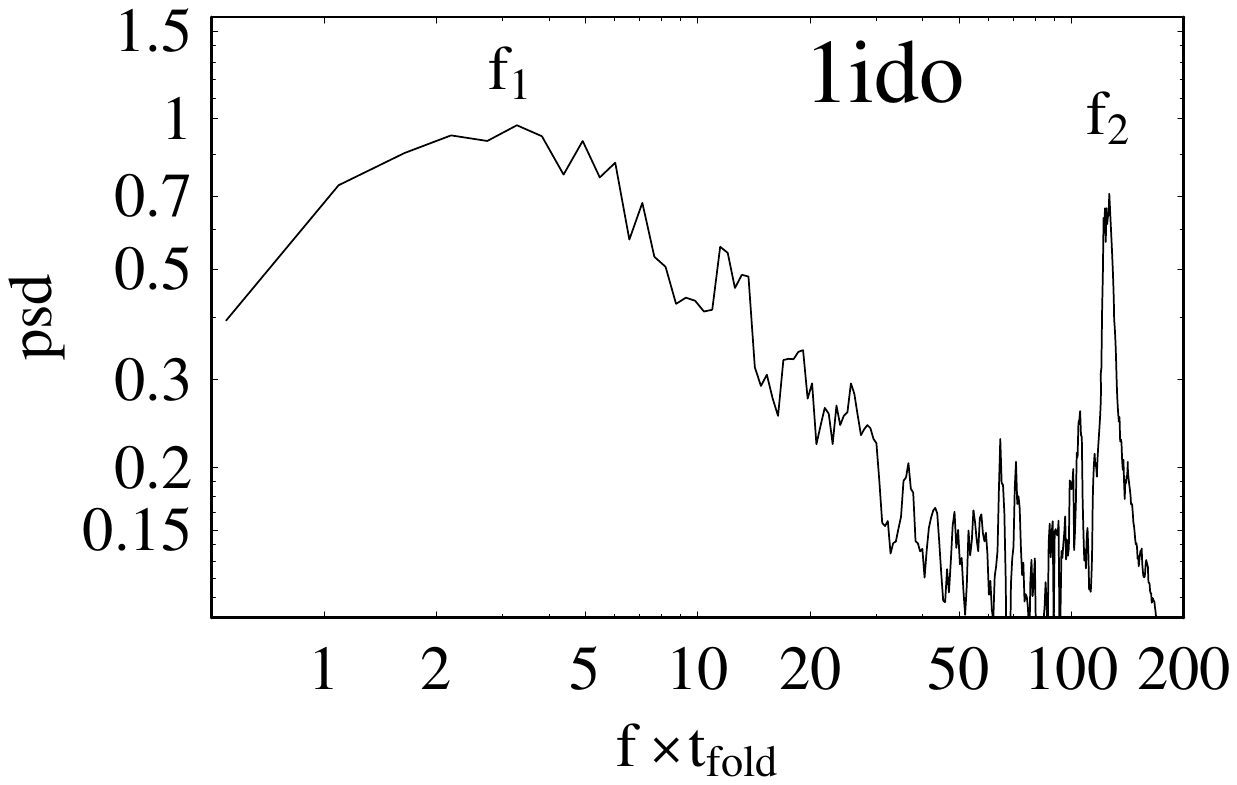}
\vspace{-5cm}
\caption{ }
\label{spectrum}
\end{figure}

\newpage

\begin{figure}
\vspace{1cm}
\includegraphics[width=10cm, viewport=30 100 500 700]{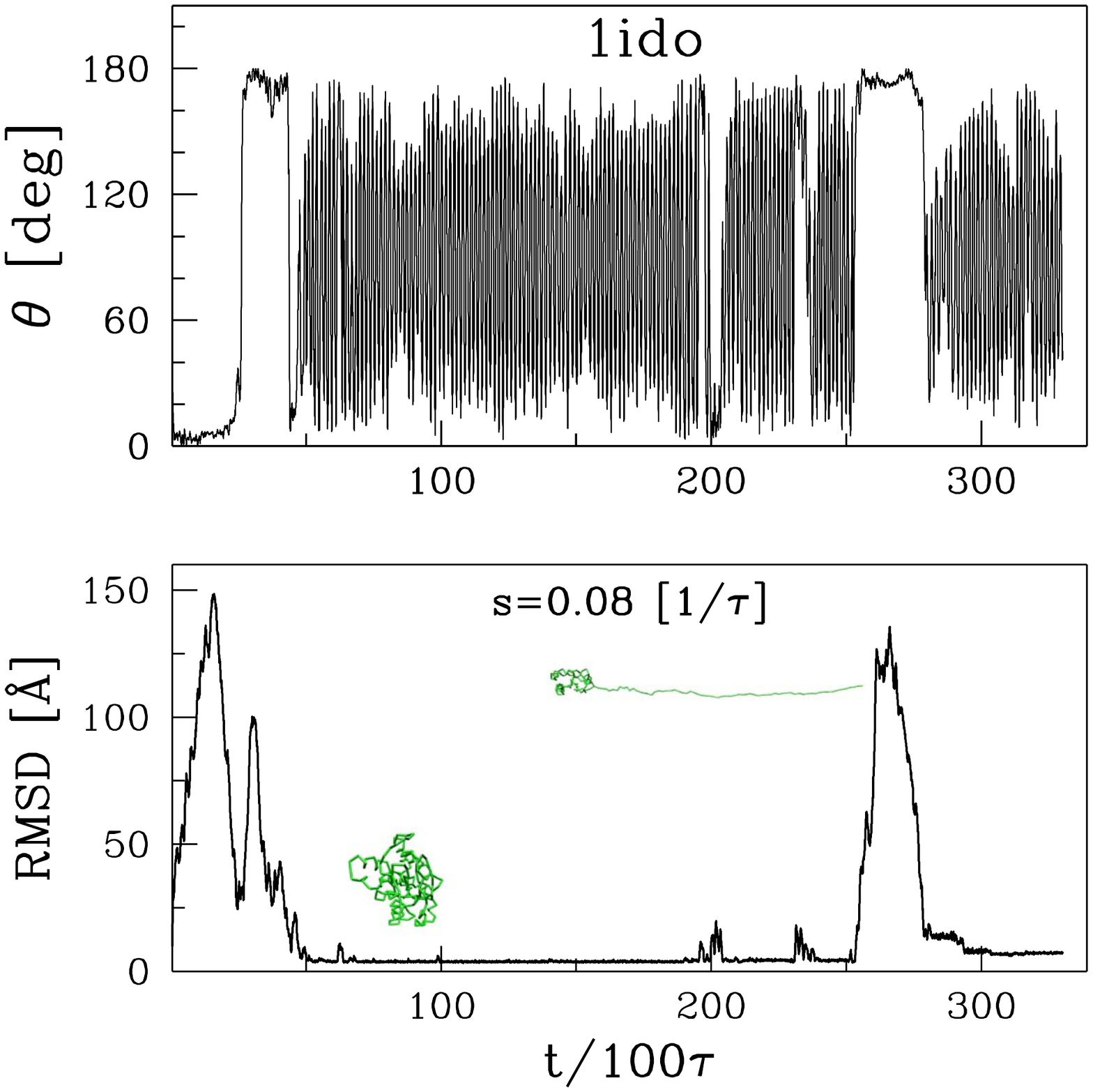}
\caption{ }
\label{angle}
\end{figure}

\newpage

\begin{figure}
\vspace{1cm}
\includegraphics[width=10cm, viewport=30 100 500 700]{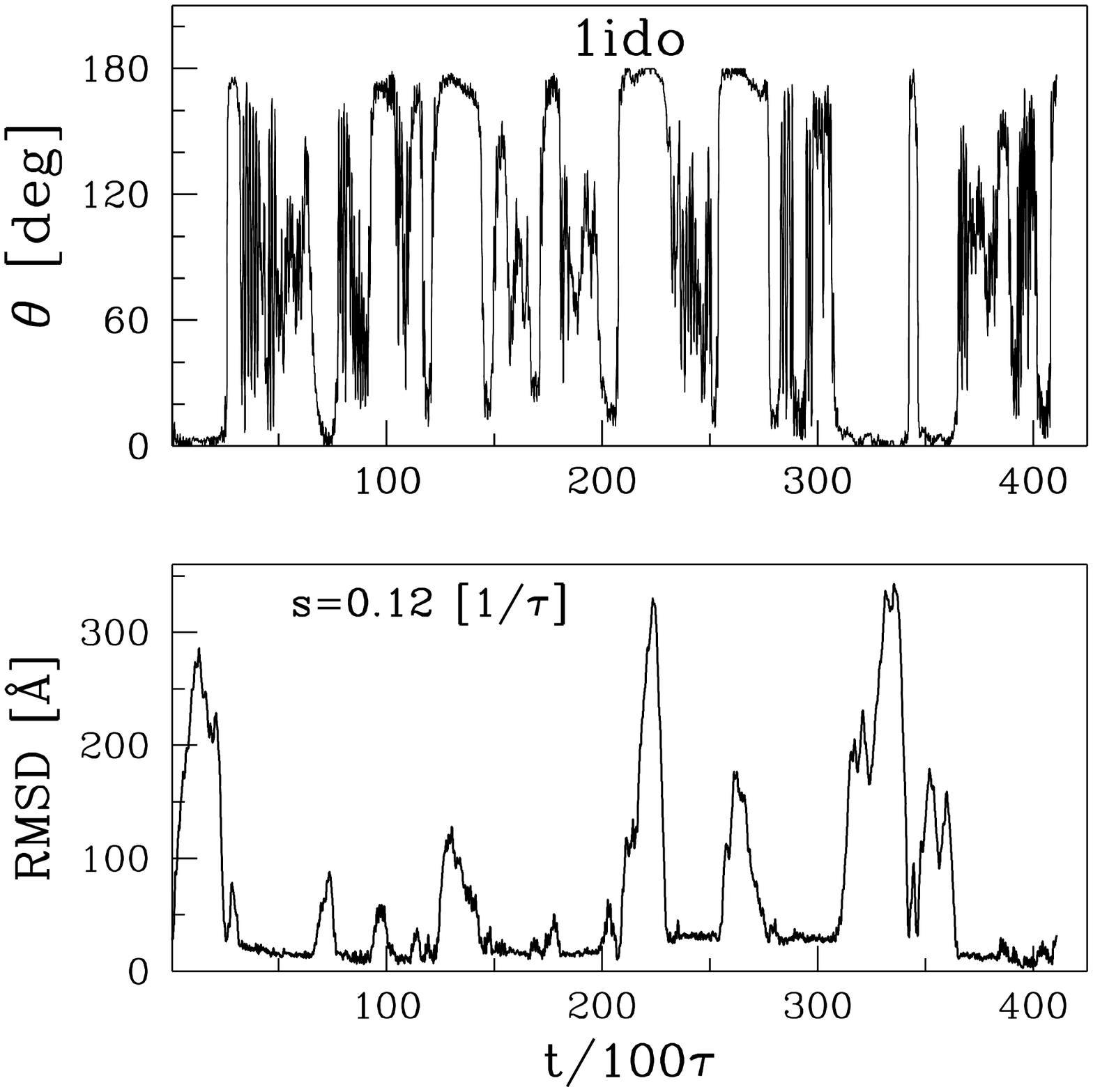}
\caption{ }
\label{anglestrong}
\end{figure}

\newpage

\begin{figure}
\vspace{1cm}
\includegraphics[width=10cm, viewport=30 100 500 700]{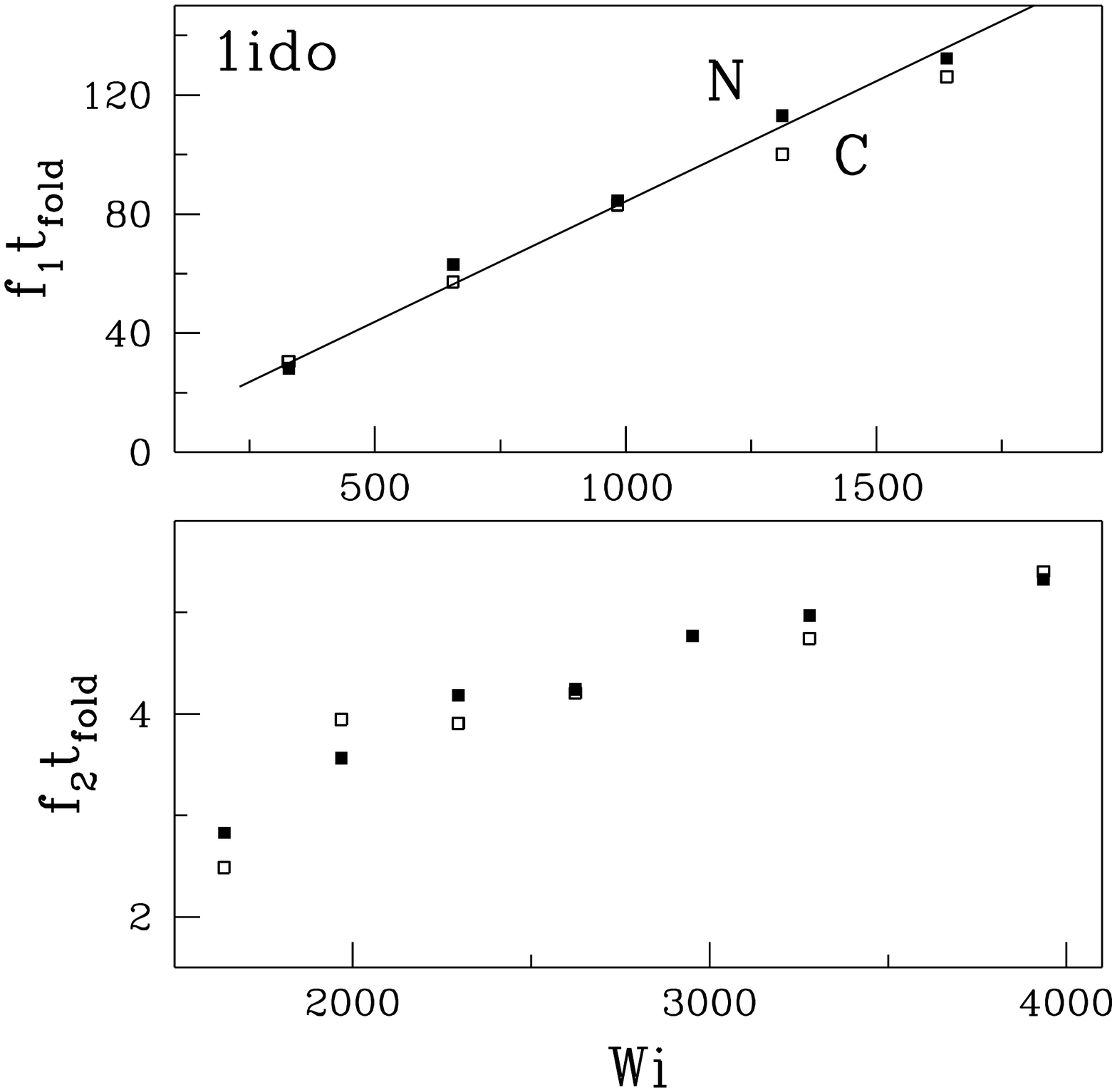}
\caption{ }
\label{freq}
\end{figure}

\newpage

\begin{figure}
\vspace{1cm}
\includegraphics[width=8cm, viewport=30 100 500 700]{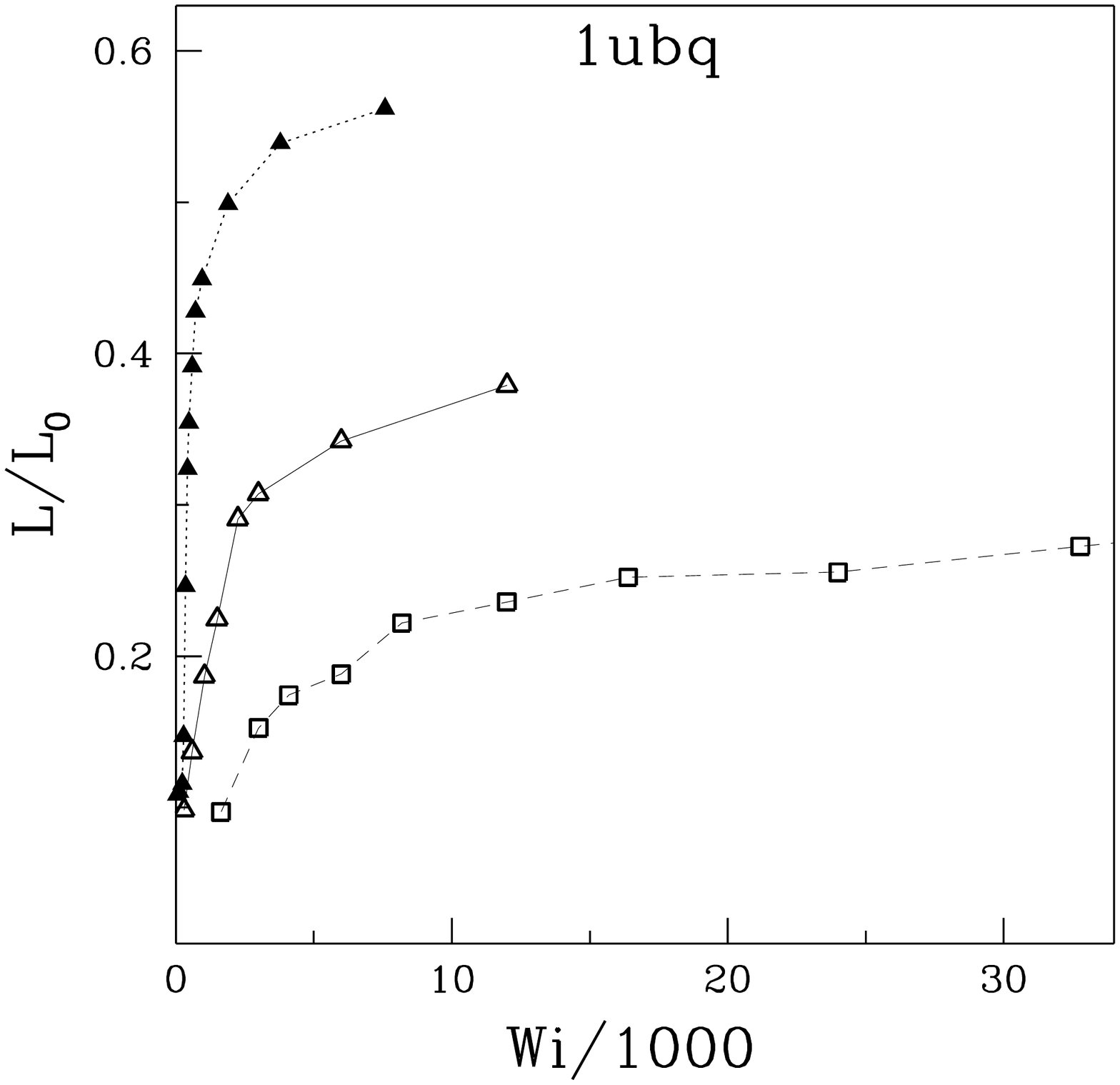}
\caption{ }
\label{ubi}
\end{figure}

\end{document}